\documentclass[12pt,preprint]{aastex}

\shorttitle{Sloan Bright Arcs Survey}
\shortauthors{Kubo et al.}

\begin{document}


\title{The Sloan Bright Arcs Survey : Six Strongly Lensed Galaxies at $\rm{z}=0.4-1.4$}

\author{Jeffrey M. Kubo\altaffilmark{1}, Sahar S. Allam\altaffilmark{1}, James Annis\altaffilmark{1}, Elizabeth J. Buckley-Geer\altaffilmark{1}, H. Thomas Diehl\altaffilmark{1}, Donna Kubik\altaffilmark{1}, Huan Lin\altaffilmark{1},  Douglas Tucker\altaffilmark{1}}
\altaffiltext{1}{Center for Particle Astrophysics, Fermi National Accelerator Laboratory, Batavia, IL 60510}


\begin{abstract}
We present new results of our program to systematically search for strongly lensed galaxies in the Sloan Digital Sky Survey (SDSS) imaging data.  In this study six strong lens systems are presented which we have \emph{confirmed} with follow-up spectroscopy and imaging using the 3.5m telescope at the Apache Point Observatory.  Preliminary mass models indicate that the lenses are group-scale systems with velocity dispersions ranging from $466-878$ km $\rm{s^{-1}}$ at $z=0.17-0.45$ which are strongly lensing source galaxies at $z=0.4-1.4$.  Galaxy groups are a relatively new mass scale just beginning to be probed with strong lensing.  Our sample of lenses roughly doubles the confirmed number of group-scale lenses in the SDSS and complements ongoing strong lens searches in other imaging surveys such as the CFHTLS \citep{cabanac07}.  As our arcs were discovered in the SDSS imaging data they are all bright ($r\lesssim22$), making them ideally suited for detailed follow-up studies.           
\end{abstract}





\section{Introduction}

Gravitational lens systems provide the opportunity to study in detail the properties of distant galaxies through the magnification provided by lensing.  Models of these systems also provide insight into the underlying mass distribution in the foreground lens.  New samples of strong lenses have been enabled by the Sloan Digital Sky Survey \citep{york00} with systems initially discovered in the SDSS \emph{spectroscopic} data \citep{bolton06}.  Because of the nature of the selection these are smaller separation systems ($<3\arcsec$, the width of the SDSS fiber radius) consisting of a single Luminous Red Galaxy (LRG) lensing a background source galaxy.  An additional parameter space to discover strong lenses is also provided by the SDSS \emph{imaging} data, and a number of complementary programs are now underway in the SDSS to search for these systems \citep{belokurov07,estrada07,hennawi08,shin08,wen08}.  Because of the moderate imaging quality (1.4\arcsec FWHM in r) in the SDSS these systems typically have a larger lens-arc separation ($>3.0\arcsec$) which is outside of the regime probed by the SDSS spectroscopic lens survey.  

Motivated by the discovery of the 8 o'clock arc \citep{allam07}, the brightest lensed Lyman Break galaxy (LBG) currently known $(z=2.73)$, we have also initiated a program to systematically search for strong lens systems in the SDSS.  Our program is focused on searching for lensed high redshift galaxies in particular $z>2$ lensed LBG's similar to the 8 o'clock arc.  Since Fall 2006 we have been following up candidate systems with imaging and spectroscopy using the 3.5m telescope at Apache Point Observatory (APO) in New Mexico.  The first lensed system discovered in our systematic search, a lensed LBG at $z=2.0$ which we named the `Clone', was recently reported in \citet{lin08}.  In addition to these lensed high redshift galaxies we are also discovering a number of interesting lensed galaxies at lower redshift.  In this paper we present our first results on six of these lower-redshift systems which we have spectroscopically confirmed to be lensed galaxies, ranging in redshift from $z=0.4-1.4$.

This letter is organized as follows: details of initial candidate selection and follow-up imaging/spectroscopy are described in $\S \ref{sec:data}$.  Preliminary mass models of each system are presented in $\S \ref{sec:sample}$.  In $\S \ref{sec:summary}$ we summarize our work and describe future directions of our survey.  Throughout we assume a standard cosmology of $\Omega_{m}=0.3$ and $\Omega_{\Lambda}=0.7$.






\section{Data}
\label{sec:data}
\subsection{Lens Search}
\label{sec:search}
The SDSS is an 8000 square degree spectroscopic and imaging survey using a dedicated 2.5m telescope at APO \citep{york00}.  We searched for candidate strong lens systems using two systematic searches on the SDSS Data Release Five (DR5) \citep{adelman07}. 

Our first candidate list was generated using the SDSS Catalog Archive Server (CAS) database.  As described in \citet{kubik07} and \citet{lin08} we performed two separate queries which searched for blue objects ($g-r<1$ and $r-i<1$) around a catalog of 221,000 Luminous Red Galaxies (LRG's) \citep{eisenstein01} and around a catalog of $29,000$ Bright Cluster Galaxies (BCG's) based on the maxBCG cluster finder algorithm \citep{koester07}.  We separated systems into groups depending on the number of blue objects found around each LRG or BCG where the number of blue objects, \emph{n}, was 2, 3, or $\geq4$.  Each list was visually inspected by four separate inspectors.  Based on their morphology, candidates were then ranked on their likelihood of being an arc.  The most promising systems in this search were chosen for follow-up.  Three confirmed systems discovered in this search are presented in $\S \ref{sec:sample}$.

Our second list of candidates is generated from a catalog of merging galaxies using the method described in \citet{allam04}.  This method was used in the initial discovery of the 8 o'clock arc \citep{allam07}.  In this technique a candidate merging galaxy pair is defined as two galaxies in the magnitude range $16.0<g<21.0$ separated by less than the sum of their respective Petrosian radii \citep{stoughton02}.   This algorithm was run on imaging data from the SDSS DR6 and the resulting catalog was visually examined and classified.  Lensing candidates were flagged by the algorithm if their morpholgies and colors were similar to a strong lens.  Three of these systems have been spectroscopically confirmed to be strong lens systems and are discussed in $\S \ref{sec:sample}$.

\subsection{APO Imaging and Spectroscopy}

We followed-up candidates with both imaging and spectroscopy using the APO 3.5m telescope.  Imaging is obtained with the SPICAM CCD imager which has a field of view of $4.8\arcmin\times4.8\arcmin$ and a plate scale of $0.28\arcsec$ per pixel.  Each target in our program is imaged in three SDSS filters $gri$ each with $3\times300s$ dithered exposures.  As described in \citet{lin08} images are reduced using standard IRAF tasks, and are co-added using the SWarp package \citep{bertin06}.  The SPICAM images are photometrically calibrated by matching unsaturated stars in the co-added images to the photometry in the SDSS.

Follow-up spectroscopy of candidate arcs was performed with the Dual Imaging Spectrograph (DIS III), a medium dispersion double spectrograph which has separate red and blue channels.  We used the standard medium red/low blue grating setup which covers a spectral range of $\rm{3600-9600}\AA$ with a resolution of $\rm{2.43\AA}$ per pixel in the blue and $\rm{2.26\AA}$ per pixel in the red.  Each arc in our sample was targeted with an exposure time of $3\times900s$.  We used a $1.5\arcsec$ or $2.0\arcsec$ slit which was typically oriented along the brightest segment of each arc in order to maximize the signal-to-noise ratio.  Spectra were also reduced using standard IRAF tasks, described further in \citet{lin08}.  Dates for all of our follow-up observations with SPICAM and DIS are summarized in Table \ref{tab:log}.

\section{Lens Sample}
\label{sec:sample}
Using the search techniques described in $\S \ref{sec:search}$ we have followed-up and confirmed a sample of six strong lens systems.  In Figures \ref{fig:arcs1} and \ref{fig:arcs2} (top row) we show color composite images of these six systems derived from our SPICAM $gri$ imaging.  Most systems have only a single visible arc, with the exception of $\rm{SDSS\ J103843.59+484917.7}$ which has a visible second arc.  The lens-arc separation in the systems range from $\sim3.7\arcsec-11.5\arcsec$ indicating that the lens galaxies are not isolated galaxies.  The color SPICAM imaging reveals several nearby galaxies in most of the systems indicating that these lenses are likely in groups or poor cluster environments. 

Unfortunately the resolution of our SPICAM imaging  is relatively poor $\sim 1.3-1.6\arcsec$ so detailed models of the systems are not possible.  Instead we adopt a simple model typically used to measure cluster scale lenses \citep{gladders02} where we assume that each system can be described by a singular isothermal sphere (SIS).  The Einstein radius of an SIS is given by 
\begin{equation}\theta_{E}=4\pi\frac{\sigma^{2}}{c^{2}}\frac{D_{ls}}{D_{s}}\end{equation} where $D_{ls}$ is the angular diameter distance between the lens-source and $D_{s}$ is the angular diameter distance to the source.  As in \cite{kubik07}, we estimate the Einstein radius from the measured lens-arc separation and knowing the lens and source redshift in each system we can calculate the velocity dispersion.  The resulting velocity dispersions are given in Table \ref{tab:lensmodel} which range from $\sim466-887$ $\rm{kms^{-1}}$, typical for group scale objects.  We also estimate the corresponding mass within the Einstein radius of each system from 
\begin{equation}
M(<\theta_{E})=\frac{c^2}{4G}\frac{D_{s}D_{l}}{D_{ls}}\theta_{E}^2.
\end{equation}

Below we briefly comment on each individual system, but we note here that redshifts of four of the systems rely on a single strong emission line seen in the red DIS spectrum.  In each case the line is best interpreted as [OII] 3727, whereas if the line were $\rm{H}\alpha$ or [OIII] 5007, we do not see the expected strong [OII] 3727 in the blue spectrum.  Similarly, if the line were Lyman alpha at high redshifts $z>4$, the arc would have dropped out from the g-band image due to the redshifted Lyman break.

\subsection{$\rm{SDSS\ J1038+4849}$}
\begin{figure}
\epsscale{1.0}
\plotone{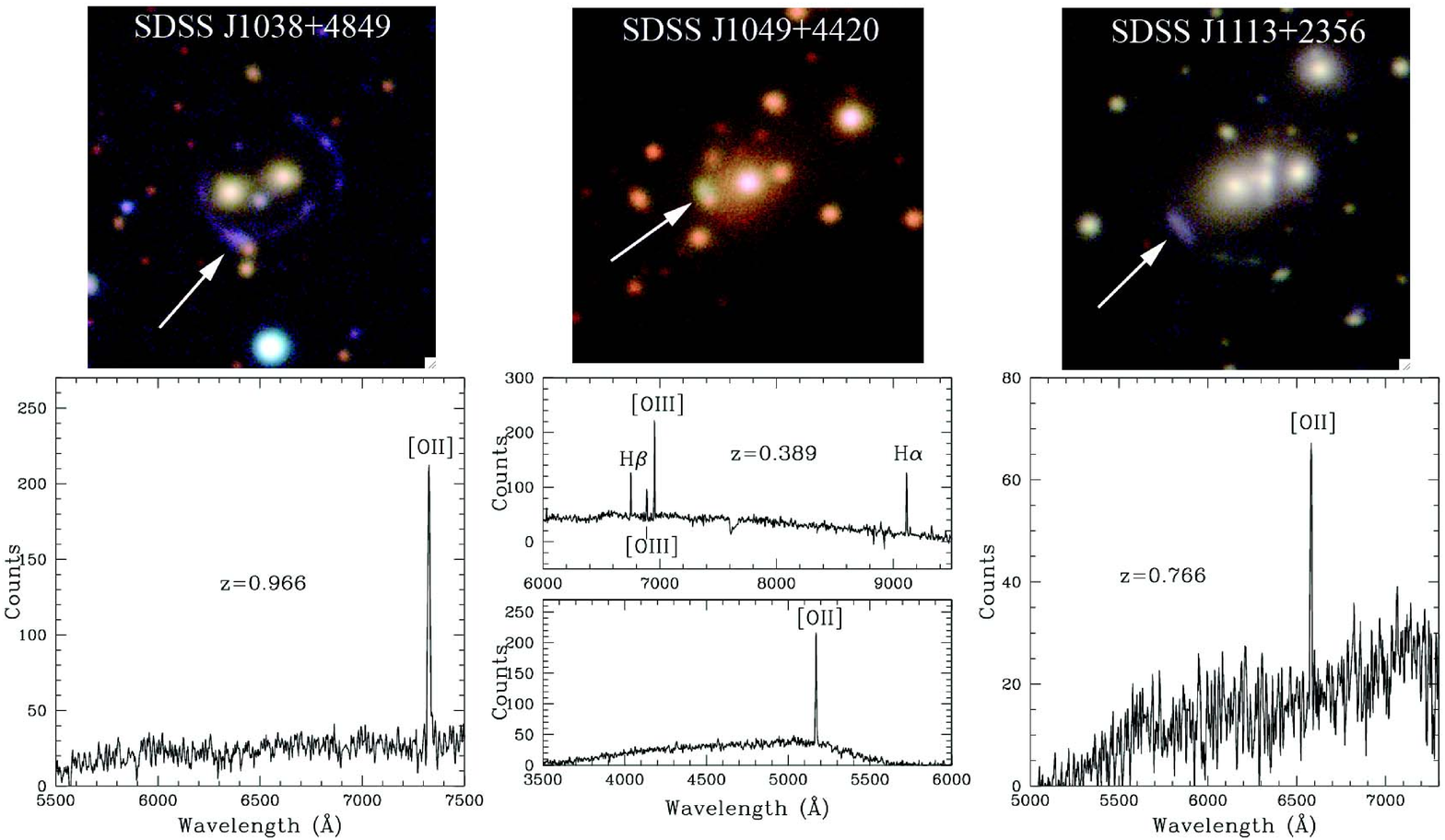}
\caption{(Top row) : Color composite images of our strong lens systems derived from our SPICAM $gri$ imaging.  In each the arrow points to the knot in the lensed background galaxy for which we obtained spectroscopy.  Images have dimensions $1\arcmin\times1\arcmin$. (Bottom row): DIS spectroscopy of the corresponding lensed source galaxy.  A strong [OII] 3727 emission line is measured in each lensed source galaxy.}
\label{fig:arcs1}
\end{figure}

$\rm{SDSS\ J1038+4849}$ has been previously reported on by \citet{belokurov07} but also independently appeared in our $n\geq4$ sample in our search around LRG's.  The system consists of two foreground LRG's, the brightest of which, SDSS J103843.58+484917.7, has a spectroscopic redshift of $z=0.4256\pm0.0003$ from the SDSS database.  The system contains two prominent arcs shown in Figure \ref{fig:arcs1}.  We independently obtained DIS spectroscopy of the bright knot along the southern arc segment (the G knot in \citet{belokurov07}) and measured the redshifted [OII] 3727 line shown in Figure \ref{fig:arcs1}.  This places the primary arc at a redshift of $z_{s}=0.966$ which is in excellent agreement with the redshift reported by \citet{belokurov07}.  We did not obtain a spectroscopic redshift of the second arc, which \citet{belokurov07} noted is likely a different lensed source galaxy.  We estimate an Einstein radius for this system of $8.4\arcsec$ which gives a velocity dispersion of $\sigma_{v}=768$ km $\rm{s^{-1}}$ and an enclosed mass of $14.0\times10^{12}\rm{M_{\odot}}$.

\subsection{$\rm{SDSS\ J1049+4420}$}
$\rm{SDSS\ J1049+4420}$ was selected from our $n=3$ sample in our search around BCG's.  The system contains one BCG with a spectroscopic redshift of $z=0.2303\pm0.0002$ from the SDSS database, with a small blue $r=21.20$ arc ($\rm{SDSS\ J104943.82+442034.2}$) to the East (Figure \ref{fig:arcs1}).  In our follow-up DIS spectroscopy of the blue arc we identify five strong emission lines, specifically [OII] 3727, $\rm{H}\beta$, [OIII] 4959, [OIII] 5007, and $\rm{H}\alpha$ (Figure \ref{fig:arcs1}) from which we obtain a secure redshift of $z_{s}=0.389$.  The relative strength of the [OIII] doublet is in the expected 1:3 ratio.  The source galaxy is the lowest redshift arc in our entire sample, but the lensing hypothesis is still valid.  We estimate an Einstein radius of $7.2\arcsec$ for this system, which gives a velocity dispersion of $805$ km $\rm{s^{-1}}$ and an enclosed mass of $8.8\times10^{12}\rm{M_{\odot}}$.

\subsection{$\rm{SDSS\ J1113+2356}$}
$\rm{SDSS\ J1113+2356}$ was identified from the merging galaxy catalog described in $\S \ref{sec:search}$ and is a small group consisting of one LRG and two bright red galaxies ($r=16.46,20.57,18.31$ from East to West) with a bright $12.6\arcsec$ long arc to the southeast (Figure \ref{fig:arcs1}).  The SDSS database places the redshift of the LRG at $z=0.3361\pm0.0002$ (Table \ref{tab:lensmodel}).  Redshifts of the two red galaxies are unknown but have the same color as the LRG and are therefore likely at the same redshift.  For spectroscopic confirmation we oriented the slit along the brightest segment of the arc and measure redshifted [OII] 3727 emission line in the DIS spectrum (Figure \ref{fig:arcs1}).  This places the redshift of arc at $z_{s}=0.766$, confirming this as a lens system.  The deeper SPICAM imaging also reveals a possible fainter extension of the arc to the West but spectra of this portion of the arc was not measured.  The measured Einstein angle of the system is $\theta_{E}=11.5\arcsec$ which corresponds to a $\sigma_{v}=882$ km $\rm{s^{-1}}$ and an enclosed mass of $21.7\times10^{12}\rm{M_{\odot}}$.   

\subsection{$\rm{SDSS\ J1137+4936}$}
\begin{figure}
\epsscale{1.0}
\plotone{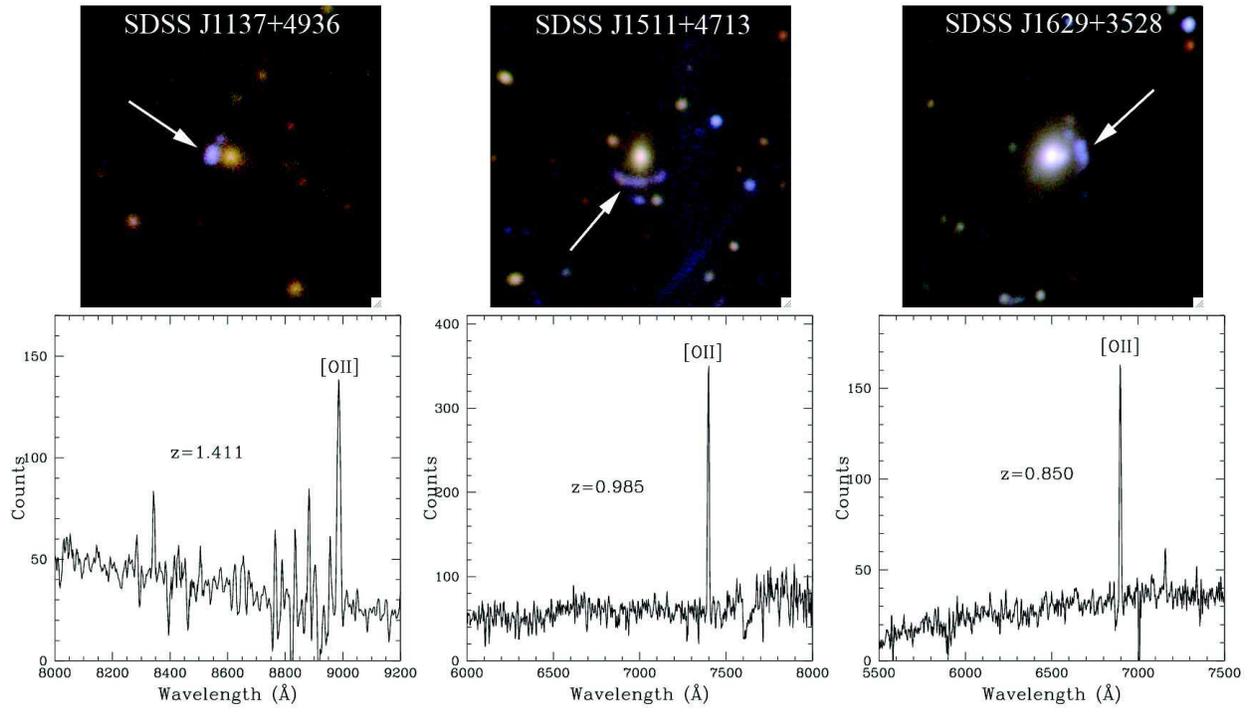}
\caption{Same as in Figure \ref{fig:arcs1}.  Additional spectral features for $\rm{SDSS\ J1137+4936}$ are shown in Figure \ref{fig:lines}.}
\label{fig:arcs2}
\end{figure}

$\rm{SDSS\ J1137+4936}$ was also selected from our merging galaxy catalog.  The system is comprised of a single LRG at $z=0.4483\pm0.0002$ with a blue arc located to the East.  In the SDSS imaging the arc is split into two knots, the brightest $(\rm{SDSS\ J113740.44+493635.8})$ with $r=20.43$ and a fainter knot $(\rm{SDSS\ J113740.26+493638.9})$ with $r=22.33$ (Figure \ref{fig:arcs2}).  The fainter knot is misclassified by the SDSS database as a star.  DIS spectroscopy of the source was measured along the bright knot and we obtain a redshift $z = 1.411$ based on a strong [OII] 3727 emission line (Figure \ref{fig:arcs2}).  This single emission line is corroborated by a number of Mg and Fe absorption features commonly observed in the optical spectra of galaxies at similar redshifts, e.g., see Fig. 9 of \citet{abraham04}.  As shown in Figure 3, we see strong absorption due to MgII 2798, MgII 2803, MgI 2852, FeII 2586, and FeII 2600, as well as weaker absorption from FeII 2344, FeII 2374, and FeII 2382, all at $z = 1.41$, consistent with the [OII] 3727 redshift.  In addition, we also see strong absorption from FeII 2586,2600 and MgII 2798, 2803 from an apparent second source galaxy at $z=1.38$.  As far as we can tell from our DIS spectroscopy (done under $2\arcsec$ seeing), there is no spatial separation between the $z= 1.41$ and $z=1.38$ absorption features, and we will need spectroscopy under better seeing conditions in order to improve the spatial resolution.  Given the present data, we interpret this system as two sources, with redshifts $z = 1.411$ and $z = 1.38$, both being lensed by a foreground LRG at $z=0.4483$.  As the bulk of the spectroscopic features are at $z=1.411$, we adopt that value in the velocity dispersion calculation.  For this system we measure an Einstein radius of $3.7\arcsec$, which gives a corresponding $\sigma_{v}=464$ km $\rm{s^{-1}}$ and an enclosed mass of $2.3\times10^{12}\rm{M_{\odot}}$.

\subsection{$\rm{SDSS\ J1511+4713}$}
$\rm{SDSS\ J1511+4713}$ was discovered in our search around SDSS LRG's and fell into our $n=3$ sample.  It consists of a foreground LRG with a spectroscopic redshift from the SDSS database of $z=0.4517\pm0.0003$, with an extended blue arc ($\rm{SDSS\ J1511+4713}$) directed to the South (Figure \ref{fig:arcs2}).  The arc contains three bright knots having a total estimated length of $10.1\arcsec$.  For spectroscopic follow-up we oriented the slit along the brightest segment of the arc and measured the redshifted [OII] 3727 emission line (Figure \ref{fig:arcs2}).  This places the arc at a redshift of $z=0.985$, confirming this as a lens system.  We estimate the Einstein radius for the system to be $5.4\arcsec$ which gives a velocity dispersion of $\sigma_{\rm{v}}=631$ km $\rm{s^{-1}}$ assuming an SIS model.  This corresponds to an enclosed mass of $6.3\times10^{12}\rm{M_{\odot}}$.

\subsection{$\rm{SDSS\ J1629+3528}$}
Our final system, $\rm{SDSS\ J1629+3528}$ was also selected from our merging galaxy catalog and contains one primary LRG with a bright (r=19.09) blue arc to the West.  A spectroscopic redshift for the LRG is not available in the SDSS database as its magnitude ($r=19.09$) is too faint to be included in the main spectroscopic survey. We have obtained DIS spectroscopy of the LRG showing strong absorption features that yield a redshift of $z=0.170$ (not shown).  For the blue arc we measure the redshifted [OII] 3727 emission line shown in Figure \ref{fig:arcs2}, which places the arc at a redshift $z_{s}=0.850$, confirming this as a lens system.  From our estimate of the Einstein radius ($5.8\arcsec$) this gives a velocity dispersion of $\sigma_{v}=510$ km $\rm{s^{-1}}$ and an enclosed mass of $2.1\times10^{12}\rm{M_{\odot}}$.  We note that the SDSS imaging also indicates a small companion galaxy which is embedded within the brighter LRG, however at our current imaging resolution we cannot estimate its magnitude.

\section{Summary}
\label{sec:summary}
We have presented six confirmed group-scale strong lenses from our search program in the SDSS. Although the focus of our survey is to confirm high redshift $(z=2\sim3)$ lensed galaxies similar to the 8 o'clock arc \citep{allam07} and the Clone \citep{lin08} we are discovering a number of lensed galaxies lying at lower redshift.  These systems were initially discovered via two systematic searches of the SDSS DR5 database and from a separate search of a catalog of candidate merging galaxies.  The arcs in our confirmed systems are relatively bright $r<22$, making them ideal candidates for detailed follow-up studies.  

Our initial follow-up APO SPICAM imaging has relatively poor resolution but preliminary modeling indicates that these systems are group-scale lenses, an intermediate regime between isolated galaxies and galaxy clusters.  This is a relatively new regime that imaging surveys are just beginning to probe with strong lensing \citep{cabanac07}.  Additional high resolution imaging will allow for more detailed models of these systems which will yield precise measurements of the mass distribution in the lens and detailed studies of the source galaxy in each system.  Future papers in this series will describe our sample of confirmed lensed $z=2-3$ galaxies as well as the results of our ongoing follow-up programs using HST, Spitzer, and other telescopes.

\acknowledgments

Fermilab is operated by the Fermi Research Alliance, LLC under Contract No. DE-AC02-07CH11359 with the United States Department of Energy.  These results are based on observations obtained with the Apache Point Observatory 3.5-meter telescope, which is owned and operated by the Astrophysical Research Consortium.  Funding for the SDSS and SDSS-II has been provided by the Alfred P. Sloan Foundation, the Participating Institutions, the National Science Foundation, the U.S. Department of Energy, the National Aeronautics and Space Administration, the Japanese Monbukagakusho, the Max Planck Society, and the Higher Education Funding Council for England. The SDSS Web Site is http://www.sdss.org/.

\clearpage



\begin{figure}
\epsscale{1.0}
\plotone{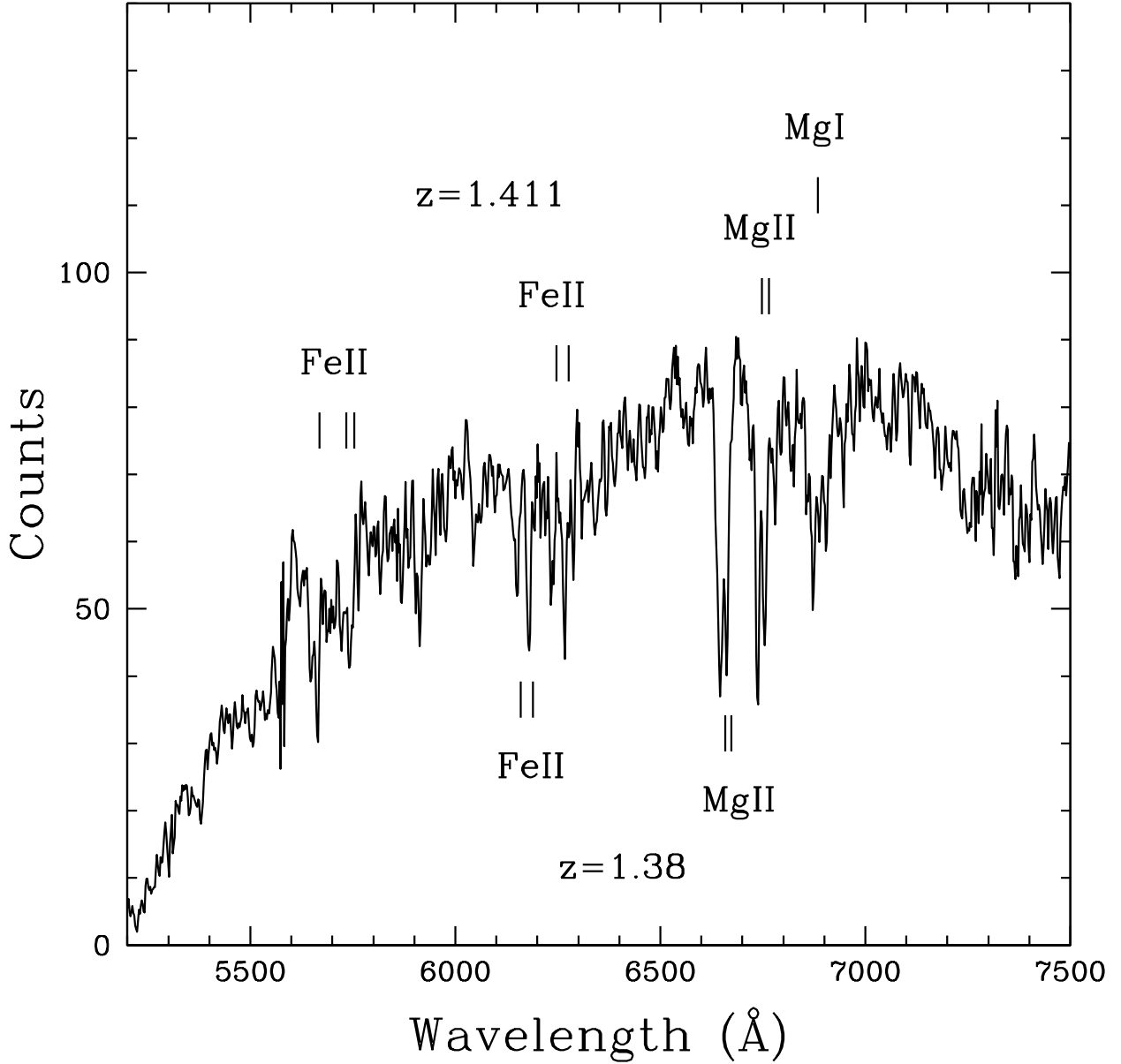}
\caption{Additional spectra lines of the lensed galaxy in $\rm{SDSS\ J1137+4936}$.  Strong absorption due to MgII 2798, MgII 2803, MgI 2852, FeII 2586, and FeII 2600 are seen, as well as weaker absorption from FeII 2344, FeII 2374, and FeII 2382, all at $z=1.41$.  This is consistent with the [OII] 3727 redshift in Figure \ref{fig:arcs2}.  Strong absorption from FeII 2586,2600 and MgII 2798,2803 from an apparent second source galaxy at $z=1.38$ is also seen. With our current level of data our interpretation is that there are two source galaxies with redshifts $z=1.411$ and $z=1.38$, being lensed by a foreground LRG at $z=0.4483$.}
\label{fig:lines}
\end{figure}

\begin{deluxetable}{ccc}
\tablecolumns{3} 
\tablewidth{0pc}
\tablecaption{Observing Log} 
\tablehead{ 
\colhead{System} & \colhead{SPICAM Imaging} & \colhead{DIS Spectroscopy}\\
}
\startdata
$\rm{SDSS\ J1038+4849}$ & 2008 Jan 3 & 2008 Feb 8\\
$\rm{SDSS\ J1049+4420}$ & 2008 Jan 3 & 2007 Mar 19\\
$\rm{SDSS\ J1113+2356}$ & 2008 Jan 3 & 2007 Nov 14\\
$\rm{SDSS\ J1137+4936}$ & 2008 Apr 5 & 2008 May 27\\
$\rm{SDSS\ J1511+4713}$ & 2008 Jun 11 & 2007 Mar 19\\
$\rm{SDSS\ J1629+3528}$ & 2008 Jun 11 & 2008 Apr 30\\
\enddata 
\label{tab:log}
\end{deluxetable} 

\begin{deluxetable}{cccccccccc}
\tablecolumns{10}
\rotate 
\tablewidth{0pc}
\tablecaption{Lens Parameters for Each System} 
\tablehead{
\colhead{System} & \colhead{R.A.} & \colhead{Decl.} & \colhead{$z_{l}$} & \colhead{$z_{s}$\tablenotemark{b}}& \colhead{$\theta_{E}$} & \colhead{$\sigma_{v}$}  & \colhead{$M(<\theta_{E})$} & \colhead{\rm{length}} & \colhead{$(g,r,i)_{\rm{lens}}$\tablenotemark{c}}\\
\colhead{} & \colhead{(deg)} & \colhead{(deg)} & \colhead{} & \colhead{}& \colhead{(\arcsec)} & \colhead{(km $\rm{s^{-1}}$)}  & \colhead{($10^{12}\rm{h^{-1}}\rm{M_{\odot}}$)} & \colhead{(\arcsec)} & \colhead{}\\
}
\startdata 
$\rm{SDSS\ J1038+4849}$ & 159.6816 & 48.8216 & $0.4256$\tablenotemark{a} & 0.966 & 8.4 & 768 & 14.0 & 16.6 & (20.33,18.59,17.87)\\
$\rm{SDSS\ J1049+4420}$ & 162.4298 & 44.3432 & $0.2303$\tablenotemark{a} & 0.389 & 7.2 & 805 & 8.8 & 3.6 & (18.56,16.96,16.47)\\
$\rm{SDSS\ J1113+2356}$ & 168.2944 & 23.9443 & $0.3361$\tablenotemark{a} & 0.766 & 11.5 & 882 & 21.7 & 12.6 & (18.18.16.46.15.83)\\
$\rm{SDSS\ J1137+4936}$ & 174.4169 & 49.6099 & $0.4483$\tablenotemark{a} & 1.411 & 3.7 & 464 & 2.3 & 5.3 & (21.18,19.38,18.64)\\ 
$\rm{SDSS\ J1511+4713}$ & 227.8281 & 47.2279 & $0.4517$\tablenotemark{a} & 0.985 & 5.4  & 631 & 6.3 & 10.1 & (20.12,18.37,17.55)\\
$\rm{SDSS\ J1629+3528}$ & 247.4773 & 35.4776 & $0.17\tablenotemark{b}$ & 0.850 & 5.8 & 510 & 2.3 & 3.0 & (18.15,16.91,16.42)\\
\enddata 
\tablenotetext{a}{Spectroscopic redshift from the SDSS database}
\tablenotetext{b}{Spectroscopic redshift determined using DIS on the APO 3.5m}
\tablenotetext{c}{Galaxy model magnitude \citep{stoughton02} from the SDSS database}
\label{tab:lensmodel}
\end{deluxetable}








\clearpage



\clearpage



\clearpage


\end{document}